 \documentstyle[twoside,fleqn,espcrc2,epsfig]{article}

\def\lab#1      {\hbox{\small #1} }
\newcommand{\be}{\begin{eqnarray}}
\newcommand{\ee}{\end{eqnarray}}
\newcommand{\ben}{\begin{eqnarray*}}
\newcommand{\een}{\end{eqnarray*}}
\newcommand{\la}{\langle}
\newcommand{\ra}{\rangle}
\newcommand{\half}{\frac{1}{2}}

\def\mb#1         {\mbox{\boldmath $#1$}}
\def\diffn#1	  {\Delta^{-}_{#1}}

\newcommand{\AmS}{{\protect\the\textfont2
  A\kern-.1667em\lower.5ex\hbox{M}\kern-.125emS}}

\begin{document}

\title{Dual Abrikosov vortex between confined charges
\thanks{Presented by R. Haymaker.  Supported in part by U.S. Dept.
                of Energy grant DE-FG05-91 ER 40617.}
}
\author{
Srinath Cheluvaraja
\address{
Department of Physics and Astronomy, 
Louisiana State University,
Baton Rouge, Louisiana, 70803 USA
},
Richard W. Haymaker$^{\rm ~a}$ and
Takayuki Matsuki
\address{Tokyo Kasei University, 1-18-1 Kaga, Itabashi, Tokyo 173-8602, Japan 
}
}
\begin{abstract}

We show that the dual Abrikosov vortex between quark and antiquark in Abelian Projected SU(2) gauge theory
is insensitive to truncation of all loops except the large monopole cluster noted by Hart and Teper. As the
transverse distance increases the discrepancy decreases, suggesting that the London penetration
depth determined by tail is invariant under the truncation of short loops.

\vspace{1pc}
\end{abstract}

\maketitle

\section{Introduction}

In 1992 the LSU group\cite{sbh}demonstrated the viability of
measuring the profile of the dual Abrikosov vortex between quark and antiquark 
as a signal
of the spontaneous U(1) gauge symmetry breaking and hence confinement 
in Abelian projected SU(2) gauge theory.  More recently Bali et.al.\cite{bss}
did a large-scale simulation confirming the picture to a higher resolution.
Gubarev et.al.\cite{gips} improved the fit to the same lattice data by
employing  a lattice Ginzburg-Landau-Higgs model rather than the continuum
version used in prior fits.  

In addition to the encouragement from the above mentioned improved lattice 
results we are interested in revisiting this study for a number of reasons.   
(i) In a recent work by DiCecio, Hart and Haymaker\cite{dhh} using a Ward Identity
 we were able to define a conserved U(1) current and hence a precise definition
of field strength and charge density.  This particular definition has not been used
prior to this in determining the profile of the vortex.  
(ii)  The Abrikosov vortex is an explicit consequence of spontaneous U(1) breaking of the vacuum\cite{pisa}
and provides a connection between this and other confinement studies.  (iii) Hart and Teper\cite{ht} have shown that 
monopoles in Abelian projected SU(2) theory fall into two groups: there is one 
very large percolating cluster of connected monopole loops and the remaining monopoles 
form much smaller clusters or small simple loops.  They showed that truncating 
all but the percolating cluster preserves the string tension.  We verify here
that this picture is preserved under the same truncation. (iv) Greensite
et.al.\cite{greensite} have noted connections between Z(2) vortex 
dynamical variables and U(1) monopoles and this approach provides an added opportunity
to further these connections.

\section{Review of Ginzburg-Landau-Higgs effective theory for Modeling of lattice data}
Consider the classical lattice field strength (in lattice units).
\ben
\widehat{F}_{\mu \nu}(\mb{m} ) 
 =
\sin \theta_{\mu \nu}.
\een
We consider a constrained Higgs field
\ben
\Phi(\mb{m} ) &=& v e^{i\chi(\mb{m} )}, \quad \quad v  = 1.
\een
Under these conditions the electric current is
\ben
\widehat{J}^e_\mu (\mb{m} ) = 
\sin
\left\{
\theta_\mu(\mb{m} )+ \chi_\mu(\mb{m} + \mu) -\chi_\mu(\mb{m} ) 
\right\}.  
\een
For small $\theta$ we get the London relations
\ben
\widehat{\cal F}_{\mu \nu}(\mb{m} ) 
\equiv
\widehat{F}_{\mu \nu}(\mb{m} ) 
- \Delta^+_\mu \widehat{J}^e_\nu (\mb{m} ) + \Delta^+_\nu \widehat{J}^e_\mu (\mb{m} )  = 0.
\een
For small $\theta$ mod $2\pi$ at the origin we obtain
a vortex with $N$ units of flux.
\ben
\widehat{\cal F}_{\mu \nu}(\mb{m} ) 
= 2 \pi N  \delta_{m_1, 0}\delta_{m_2, 0}.
\een
This assumes an infinite Higgs mass $M_H$. With a finite mass 
there is a transition region of size $\sim 1/M_H$ in the core of the vortex but the
above London relation holds outside the core. 
This suggests that we
look for this relation far from the source.  The quarks need not be
far apart to check this.

\section{Precise lattice Abelian flux}
We can define the classical lattice 
field strength and conserved current through the
relation
\ben
\Delta_{\mu} F_{\mu \nu} &=& J_{\nu},
\een
\ben
0 = \Delta_{\mu}\Delta_{\nu} F_{\mu \nu} &=& \Delta_{\nu} J_{\nu}.  
\een
Zach et.al.\cite{zfks} noted that these relations can be derived from a
Ward identity for lattice averages of the U(1) gauge theory 
\ben
 ea^2F_{\mu \nu}  &=&
   \frac{\la \sin \theta_{\mu \nu}  \sin \theta_{W} \ra}{\la \cos \theta_W \ra}.
\een
\ben
J_\mu   &=&  J_\mu^{\lab{ext.} }.  
 \een
Evaluating the divergence of the electric field on a time-like line of a Wilson loop
gives
\ben
\left.\Delta \cdot (ea^2 \mb{E} )\right|_{\mb{n}= \mb{n} _0} &=& e^2 = \frac{1}{\beta_{U(1)}},
\een
and zero otherwise. 

In the generalization to Abelian projected SU(2) in the maximal Abelian gauge,
multiplication of a link by a group element requires an associated 
gauge transformation to maintain the gauge condition
\cite{dhh}.  Taken together we obtain  
\ben
ea^2F_{\mu \nu}   &=& 
 \frac{\la \half \mb{tr} (i\sigma_3 D _{\mu \nu})  \sin \theta_{W} \ra}{\la \cos \theta_W \ra},
\een
where the link variable is separated into the diagonal and off-diagonal parts
\ben
U_{\mu}  &=&   D_{\mu} +   O_{\mu}, 
\een
and $D _{\mu \nu} $ is a plaquette constructed from the diagonal parts.  The subsequent
current defined by the divergence gets contributions from the external source,
the charged fields, the gauge fixing and from ghosts.
\ben
J_\mu   &=&  J_\mu^{\lab{ext.} } + J_\mu^{\lab{dyn.} } +  J_\mu^{\lab{g.f.} } + J_\mu^{\lab{ghosts} }.  
\een

Separating the diagonal parts of the links gives the photon part of the action  
\ben
 S &=& \beta \sum  \half\lab{tr} (D_{\mu \nu}) + \cdots \\
 &\approx&  \beta \sum
 \la \cos \phi \ra^4 \cos \theta_{\mu \nu}  + \cdots   \\
 &\approx&  \frac{1}{e^2} \sum
   \cos \theta_{\mu \nu}  + \cdots,
\een
where we used the fact that $\la \cos \phi \ra $ has small fluctuations 
in this gauge\cite{poulis}. 
\ben
\frac{1}{e^2}&\approx& \beta  \la \cos \phi \ra^4 =  2.5115  \times (0.9331(2))^4, \\
e^2 &\approx& 0.53. 
\een
This identifies the U(1) charge that determines the electric flux quantization  which 
will be discussed in a subsequent paper.  

The divergence of the electric current
measured on a time like Wilson line in the classical limit
gives
\ben
\left.\Delta \cdot (ea^2 \mb{E} )\right|_{\mb{n}
= \mb{n} _0} &=& \frac{1}{\beta_{SU(2)}}  = 0.40  \quad \lab{(bare)}.
\een
In the quantum case, the charged field dresses the bare charge and gives in this case
\ben
  &=& 0.51  \quad \quad   \lab{measured (dressed)},
\een
showing that there is significant screening even at the shortest distances.
\begin{figure}[h]
\begin{center}
\epsfig{file=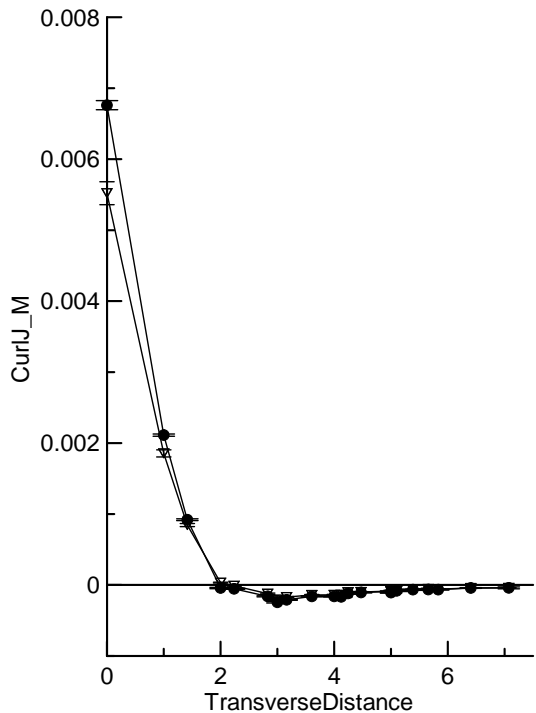, width=7.7cm}
\end{center}
\caption{Solid symbols are the curl of the monopole current vs. the transverse 
distance from the center of a Wilson loop;  $\beta = 2.5115$,
lattice $20^4$, $3 \times 3 $ loop  with fat space links. 
The open symbols are calculated from truncated monopole loops.}
\end{figure}
\begin{figure}[h]
\begin{center}
\epsfig{file=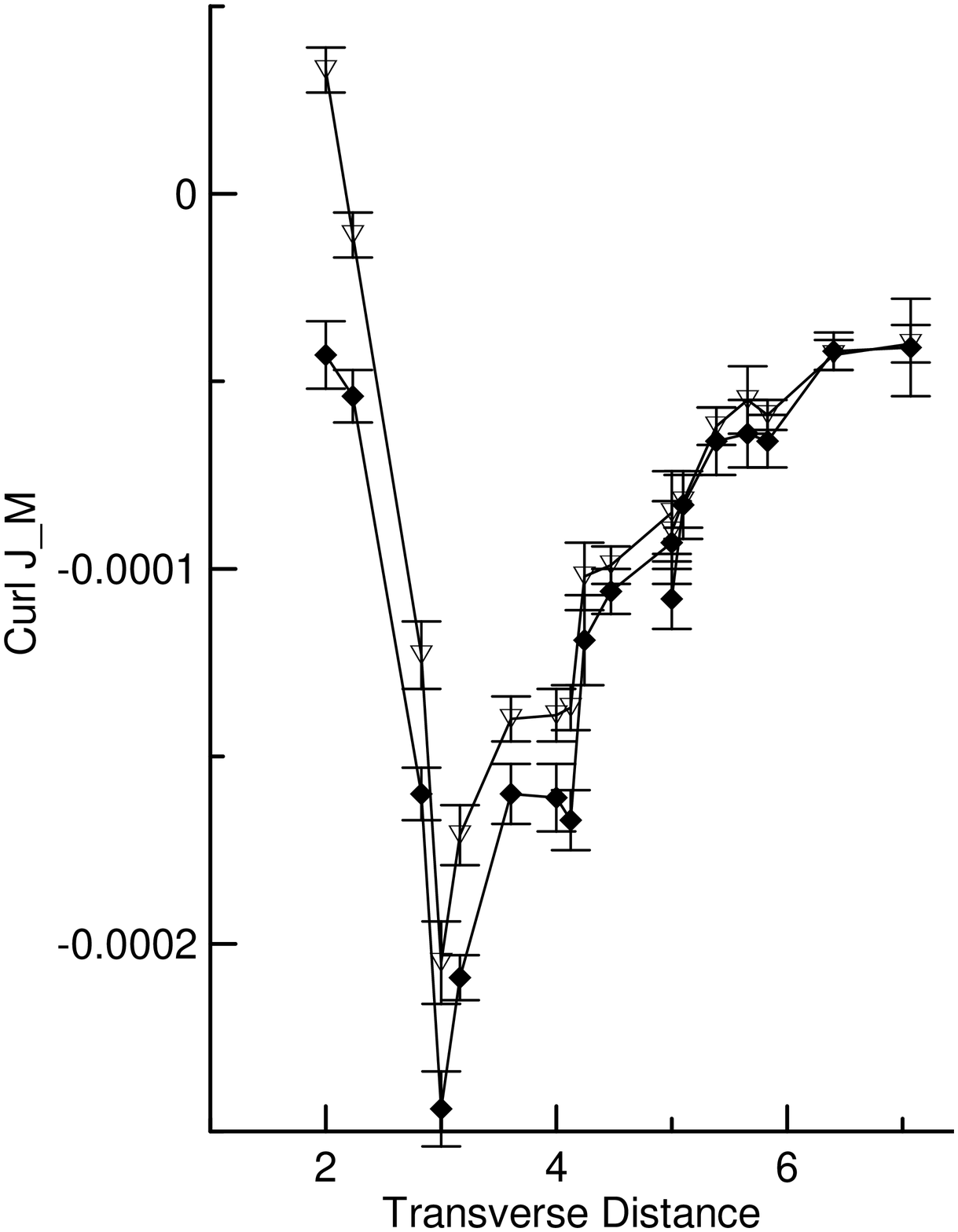, width=7.7cm}
\end{center}
\caption{Rescaled Fig. 1. Lines are not fits.}
\end{figure}
\section{Percolating monopole cluster}
Hart and Teper\cite{ht} showed that for large volumes, the monopole
currents fall into two distinct classes.  There is one large 
percolating cluster
that permeates the whole lattice volume and it gives the full
string tension.  On this cluster, scaling is observed for 
the current density and magnetic screening mass.  The remaining
loops are localized and appear to give no contribution to
the string tension.

Fig. 1 gives the profile of the curl of the monopole current
as a function of the transverse distance from the source, 
with and without the truncation of all but the percolating cluster.
\newpage
Fig. 2 shows the blowup of the tail region.  We see that 
as the transverse distance increases, the effect of the truncation 
is suppressed indicating that the London penetration depth is
due to the percolating cluster alone.

\end{document}